\documentclass[12pt]{article}

\usepackage{siunitx} 
\usepackage[margin=1in]{geometry}
\usepackage{amsmath}
\numberwithin{equation}{section}
\usepackage{graphicx}
\usepackage{natbib}
\usepackage{underscore}
\usepackage{physics}
\usepackage[ruled,vlined]{algorithm2e}
\usepackage{algpseudocode}
\usepackage{booktabs}
\usepackage{amsfonts}
\usepackage{bm}
\usepackage{threeparttable}
\usepackage{setspace}
\usepackage{rotating}
\usepackage[width=.95\textwidth, skip=0.1\baselineskip]{caption}

\allowdisplaybreaks

\usepackage[export]{adjustbox}
\usepackage[colorlinks, citecolor=blue,  urlcolor=blue,
linkcolor=blue]{hyperref}
\usepackage{xcolor}
\usepackage[normalem]{ulem}

\usepackage[]{lineno}
\newcommand*\patchAmsMathEnvironmentForLineno[1]{%
	\expandafter\let\csname old#1\expandafter\endcsname\csname
	#1\endcsname
	\expandafter\let\csname oldend#1\expandafter\endcsname\csname
	end#1\endcsname
	\renewenvironment{#1}%
	{\linenomath\csname old#1\endcsname}%
	{\csname oldend#1\endcsname\endlinenomath}}%
\newcommand*\patchBothAmsMathEnvironmentsForLineno[1]{%
	\patchAmsMathEnvironmentForLineno{#1}%
	\patchAmsMathEnvironmentForLineno{#1*}}%
\AtBeginDocument{%
	\patchBothAmsMathEnvironmentsForLineno{equation}%
	\patchBothAmsMathEnvironmentsForLineno{align}%
	\patchBothAmsMathEnvironmentsForLineno{flalign}%
	\patchBothAmsMathEnvironmentsForLineno{alignat}%
	\patchBothAmsMathEnvironmentsForLineno{gather}%
	\patchBothAmsMathEnvironmentsForLineno{multline}%
}

\allowdisplaybreaks 
\let\proglang=\textsf
\newcommand{\pkg}[1]{{\fontseries{m}\selectfont #1}}

\usepackage{multirow, makecell}
\usepackage{booktabs,array}
\usepackage[affil-it]{authblk}

\title{Recurrent Events Modeling Based on a Reflected Brownian Motion
  with Application to Hypoglycemia}

\date{}

\newcommand{\blind}{BLINDCODE}

\if1\blind
  \author{Anonymous Author}
\else
  \author[1]{Yingfa Xie}
  \author[2]{Haoda Fu}
  \author[3]{Yuan Huang}
  \author[1]{Vladimir Pozdnyakov}
  \author[1]{Jun Yan \thanks{Email address: \texttt{jun.yan@uconn.edu}; 
    corresponding author}}
  \affil[1]{Department of Statistics, University of Connecticut}
  \affil[2]{Eli Lilly and Company}
  \affil[3]{Department of Biostatistics, Yale School of Public Health}
  
\fi

\begin{document}
\maketitle

\begin{abstract}
  Patients with type 2 diabetes need to closely monitor blood sugar
  levels as their routine diabetes self-management. 
  Although many treatment agents aim to tightly control blood sugar,
  hypoglycemia often stands as an adverse event. 
  In practice, patients can observe hypoglycemic events
  more easily than hyperglycemic events due to the perception of neurogenic
  symptoms. We propose to model each patient's observed
  hypoglycemic event as a lower-boundary crossing event for a reflected
  Brownian motion with an upper reflection barrier. The lower-boundary
  is set by clinical standards. To capture patient heterogeneity and
  within-patient dependence, covariates and a patient level frailty are
  incorporated into the volatility and the upper reflection barrier.
  This framework provides quantification for the underlying glucose
  level variability, 
  patients heterogeneity, and risk factors' impact on glucose. We
  make inferences based on a Bayesian framework using
  Markov chain Monte Carlo. Two model comparison criteria, 
  the Deviance Information Criterion and 
  the Logarithm of the Pseudo-Marginal Likelihood, are used
  for model selection. The methodology is validated in
  simulation studies. In analyzing a dataset 
  from the diabetic patients in the DURABLE trial, our model provides adequate
  fit, generates data similar to the observed data, and offers insights
  that could be missed by other models.
\end{abstract}

\vfill

\doublespacing

\section{Introduction}

Diabetes is a group of disease characterized by elevated
blood glucose. It is a major cause of kidney failure, non-traumatic lower-limb
amputations, blindness, heart disease, and stroke. As a result, diabetes is one
of the leading causes of death. Diabetes affects 37.3 million Americans 
which is 11.3\% of the US population \citep{cdc2022}.
The goal of treating diabetes patients is to lower their blood glucose in
range. When blood glucose level is too high, patients will experience
\emph{hyperglycemia}. Lowering glucose too much, however, could result in an
adverse event called \emph{hypoglycemia}. Concern over hypoglycemia has a
significant negative impact on diabetes management, making it a major factor to
prevent patients’ glycemic control from reaching treatment target
\citep{wild2007critical}. It is desired to develop new anti-diabetes
agents that lead to less hypoglycemic events while lowering the glucose toward
the normal range.

Hypoglycemias are relatively easier observed than hyperglycemias in practice. 
Symptoms of hypoglycemia are the result of brain glucose deprivation or the
perception of physiological change. Awareness of hypoglycemia is mainly the
result of the perception of neurogenic symptoms
\citep{towler1993mechanism, derosa2004hypoglycemia}.
Times of hypoglycemic events are recorded through patients'
self-report diaries with easily observed symptoms, such as headache,
shaking, sweating, hunger, and fast heartbeat \citep{cryer2003hypoglycemia}.
In contrast, hyperglycemia symptoms are much less obvious, and may not be
noticed
unless a blood sugar level test is performed \citep{cryer2009evaluation}.
Therefore, only the times of recurrent hypoglycemic events are
reliably available in self-reported data. These event times have been the target
of modeling in diabetes clinical research
\citep[e.g.,][]{fu2016hypoglycemic, ma2021heterogeneous, doubleday2022risk}.

A variety of recurrent event models have been available.
Earlier works characterize the intensity function of
the recurrent event process \citep{andersen1982cox}. The multiple events can be
analyzed with the conditional intensity for the sequence of events given the
history \citep{prentice1981regression, lee1992cox} or with marginal intensities
of the gaps between successive events \citep{wei1989regression,
  huang2003marginal, schaubel2004regression}. Alternative
approaches to gap times have been proposed, such as accelerated failure time
model \citep{chang2004estimating}, additive hazards model
\citep{sun2006additive}, and quantile regression
\citep{luo2013quantile}. Heterogeneities beyond covariates can be incorporated
in the models as a subject-level random effect which also captures the
dependency between events within an individual 
\citep{klein1992semiparametric, duchateau2003evolution, box2006repeated}. 
Less restrictive model assumptions have been considered by marginal rate or
marginal mean models \citep{lawless1995some, lin2000semiparametric}, which do
not fully specify the recurrent event process. More
recently, a general scale-change model encompasses multiple existing
event time models in a unified framework and allows a flexible form of
informative censoring \citep{xu2020generalized}. The readers are referred to
recent reviews for details \citep{cook2007statistical, charles2019analyze}.

An alternative approach is the first hitting time (FHT) model. An FHT model has
three components: a stochastic process, a boundary, and a time scale on which
the process unfolds \citep{lee2019survey}. An event occurs when the underlying
stochastic process hits the preset boundary. It is well-known that the
distribution of the FHT of a Brownian motion is the inverse Gaussian
distribution \citep{schrodinger1915theorie, folks1978inverse}. Incorporating
covariates into the model parameters leads to threshold regression
\citep{lee2006threshold}. Random effects can be incorporated to account for
unmeasured covariates \citep{pennell2010bayesian}. For recurrent events,
\citet{whitmore2012recurrent} modeled the event times as a sequence of
independent and identically distributed hitting times of a Wiener process as it
passes through successive equally-spaced levels. \citet{economou2015bayesian}
extended the inverse Gaussian threshold regression model with a random effect.
\citet{malefaki2015modelling} further extended the model to allow censoring to
occur at every intermediate stage during the recurrent event process. For
recurrent hypoglycemic events of diabetic patients, it is natural to model them
as hitting times when the glucose level crosses a lower barrier but the
unobserved hyperglycemic event times need to be handled.

We propose to model hypoglycemic event times as the FHTs of a Brownian motion
with an upper reflection boundary hitting a lower barrier. The upper reflection
barrier represents that patients' glucose levels will eventually decrease,
potentially due to the use of glucose-lowering medications that are routinely
prescribed for this patient population. Consequently, hyperglycemic events at
which the patient's glucose levels hit the upper reflection barrier are
typically not observed. Conversely, the lower barrier represents a boundary
beyond which patients experience uncomfortable symptoms associated with
hypoglycemia. As a result, hitting times for the lower barrier are typically
observed. The distribution of the FHT of the lower barrier of a Brownian motion
with an upper reflection barrier has been studied by
\citet{hu2012hitting}. Taking advantage of this distribution,
we model the recurrent hypoglycemic events based on a sequence of
reflected Brownian motions hitting a lower barrier, after which the process
restarts at a preset point between the two barriers. The gap times between the
successive events of the same patient are assumed to be independent conditional on
a subject-level random effect. Covariates and a subject-level random effect are
linked to parameters of the FHT model. The model provides unique opportunities 
to characterize the variability and heterogeneity of hypoglycemic events with an
intuitive interpretation.

Replacing the Brownian motion with a reflected Brownian motion in an FHT model
introduces a significant level of complexity. The density and the distribution
functions of the FHT derived in \citet{hu2012hitting} have not been used in the
statistical literature. Both functions involve infinite series, which present
challenges to accurate evaluation of the loglikelihood and efficient design of
random number generation. To address the challenges,
we first demonstrated that the right tail of the FHT is bounded by an exponential
rate. Drawing on this result, we implemented the density and distribution
functions and formulated an efficient rejection sampling algorithm with a
three-piece proposal density. Inferences are conducted  
within a Bayesian framework using Markov chain Monte Carlo (MCMC) with our
implementation of the FHT distribution incorporated into the generic algorithm
in \proglang{R} package \pkg{NIMBLE} \citep{devalpine2017programming}. Different models,
especially those with different types of random effects, are compared with two
Bayesian model comparison criteria, deviance information criterion (DIC)
\citep{spiegelhalter2002bayesian} and logarithm of the pseudo-marginal
likelihood (LPML) \citep{geisser1979predictive, gelfand1994bayesian}.
The whole methodology was validated in numerical studies before being applied to
analyze the hypoglycemic event data in our motivating application. The
\proglang{R} code is publicly available in
\url{https://github.com/YingfaX/reflbrown}.

The remainder of the article is organized as follows.
In Section~\ref{sec:model}, an FHT model of a
reflected Brownian motion is set up and an efficient rejection sampling
algorithm is proposed for simulation from the model.
Bayesian parameter estimation and model
selection methods are presented in Section~\ref{sec:BayesEst}.
Simulation studies that validate the Bayesian inferences are reported in
Section~\ref{sec:simulation}. The methodologies are applied to the hypoglycemic
event times from the DURABLE trial (ClinicalTrials.gov Identifier: NCT00279201)
in Section~\ref{sec:aplc}. Section~\ref{sec:conclusion} concludes
with a discussion.

\section{Model} \label{sec:model}

The fundamental component of our model is the FHT model for the first time when
a Brownian motion with a reflecting upper barrier hits a lower barrier.  A
sequence of such models are assembled for recurrent events. Subsequently, we
incorporate covariates and frailty terms in the model parameters to further
characterize variability and heterogeneity.

\subsection{FHT of a reflected Brownian motion} \label{subsec:FHT}
Consider a no-drift Brownian motion~$X(t)$ with volatility~$\sigma$. Without
loss of generality, let $\kappa$ be the upper reflection barrier $\kappa > 0$
and $\nu = 0$ be the lower barrier. Suppose that 
$X(0) = x_0 \in [0, \kappa]$ is the starting point. The first
time when $X(t)$ hits~$\nu$ is $\tau := \inf \{t > 0; X(t) = \nu\}$. For any
$\nu \in [0, x_0)$, the density and 
distribution function of $\tau$ are, respectively \citep{hu2012hitting},
\begin{align}
  \label{eq:dens}
  f(t|x_0, \nu, \kappa, \sigma) &= \sum_{n=1}^{\infty} c_n \lambda_n e^{-\lambda_n t}, & t > 0, \\
  \label{eq:dist}
  F(t|x_0, \nu, \kappa, \sigma) &= 1 - \sum_{n=1}^{\infty} c_n e^{-\lambda_n t}, & t > 0,
\end{align}
where for $n = 1, 2, \ldots$,
\begin{align*}
  \lambda_n= \frac{(2n-1)^2 \sigma^2 \pi^2 }{8(\kappa - \nu)^2},
  \mbox{ and }
  c_n       = \frac{(-1)^{n + 1}4 }{(2n-1)\pi}
                \cos(\frac{ (2n - 1)\pi(\kappa - x_0)}{2(\kappa - \nu)}).
\end{align*}
Note that 
$0 < \lambda_1 < \lambda_2 < \cdots$, $\lambda_n \to \infty$, and
$\sum_{n=1}^{\infty} c_n = 1$. It is tempting to think of $f(t)$ as a mixture of
exponential densities, but this is not true because $c_n$'s can be negative.

Random number generation from density~\eqref{eq:dens} has not been investigated
in the literature, we propose a rejection sampling algorithm. By construction,
the proposal density~$g(t)$ is required such that $f(t)/g(t)$ is bounded over
the support~$t > 0$. It is known that $f(t)/t \to 0 \mbox{ as } t \downarrow 0$
\citep[p.~13]{hu2012hitting}. In Appendix~\ref{subsec:asymp}, we show that the
right tail of $f(t)$ is bounded by that of an exponential density with
rate~$\lambda_1$:
\begin{equation*}
  f(t) \sim c_1 \lambda_1 \exp(-\lambda_1 t) \mbox{ as } t \to \infty.
\end{equation*}
Given these properties, we propose a three-piece
proposal density which consists a left tail, a body, and a right tail.
The left tail is a triangular density that connects the origin to the peak of
the target density peak
$(t_m, f(t_m))$, where $t_m$ is the mode of the target density~$f$. 
The body is a trapezoid density 
that connects the peak $(t_m, f(t_m))$ to a user-defined $q^{th}$ 
quantile point $(t_q, f(t_q))$ where $t_q$ is the upper $q$th quantile of~$f$,
and $q$, for example, can be set to be 0.95. The right tail beyond $t_q$ is the
tail of the exponential density with rate~$\lambda_1$. The details of
the implementation of the algorithm is in Appendix~\ref{subsec:rejSpl}.

\subsection{Modeling gap times of hypoglycemic events}

\begin{figure}
\centering
\includegraphics{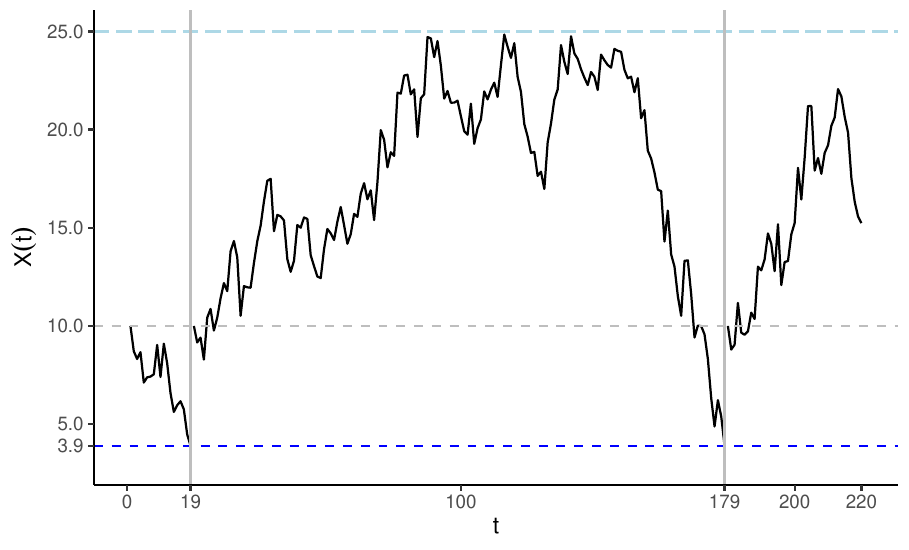}  
\caption{A sample path (black solid line) of the reflected Brownian motion with
$x_0 = 10$ (grey dash line), $\kappa = 25$ (lightblue dash line), $\nu = 3.9$
(blue dash line), and $\sigma = 3$. Two events occur at time~$T_1 = 19$ and 
$T_2 = 179$, respectively, as marked by the gray vertical lines. Observation is
censored at time~$220$.}
\label{fig:samplePathPlot}
\end{figure}

We model the recurrent hypoglycemic events based on a sequence of
reflected Brownian motions hitting a lower barrier~$\nu$. The Brownian motion
has volatility~$\sigma$ and an unobserved reflection upper barrier~$\kappa$,
beyond which hyperglycemia occurs but may not be recorded. The reflected 
Brownian motion restarts at a known starting point~$x_0$ after hitting the
lower barrier~$\nu$, where $x_0$ can be set as the average level from
the patients group. This is motivated by that patients' glucose level
arises back, possibly after eating to relieve quickly the symptoms 
of hypoglycemia. The gap times between two successive 
hypoglycemic events thus follow a FHT distribution. 
Figure~\ref{fig:samplePathPlot} shows a sample path for a subject who
experiences two events at time~$T_1 = 19$ and $T_2 = 179$, and finally is 
censored at time~$220$.

To incorporate heterogeneity, we allow volatility~$\sigma$ and upper
barrier~$\kappa$ to depend on subject-level covariates. Appropriate link
functions are necessary to ensure that $\sigma > 0$ and $\kappa > x_0$.
Subject-level random effects can be further added to the regression. 
In particular, the models for $\sigma_i$ and $\kappa_i$ with the log link are
\begin{equation*}
  \log(\sigma_i) = \bm{X}_i^{\top} \bm{\beta} + Z_{i1}, \quad\text{ and }\quad
  \log(\kappa_i - x_0) = \bm{X}_i^{\top} \bm{\alpha} + Z_{i2},
\end{equation*}
respectively, where $\bm{X}_i$ is a $p$-dimensional covariate vector for
subject~$i$, $\bm{\beta}$ and $\bm{\alpha}$ are $p$-dimensional regression
coefficient vectors, and $(Z_{i1}, Z_{i2})$ is a bivariate normal random effect
with mean $\bm{0}$, variance $\bm{\theta}= (\theta_1, \theta_2)$, and
correlation $\rho$. Also note that, $\bm{X}_i$ is considered as time-independent
covariate vector.

For ease of implementation, the bivariate 
normal vector~$(Z_{i1}, Z_{i2})$ can be reparametrized 
as~$(Z_{i1}, \gamma Z_{i1} + Z_{i2}')$,
where $Z_{i1}$ and $Z_{i2}'$ are independent normal variables
with mean zero and variance~$\theta_1$ and $\theta_2'$, respectively. 
Then the variance of $Z_{i2}$, $\theta_2$, is reparametrized as 
$\gamma^2 \theta_1 + \theta_2'$,
and the correlation~$\rho$ between $Z_{i1}$ and $Z_{i2}$ is reparametrized as 
$\gamma \sqrt{\theta_1/(\gamma^2 \theta_1 + \theta_2')}$.
The reparametrized models for $\sigma_i$ and $\kappa_i$ are, respectively,
\begin{equation}
  \label{eq:model}
  \log(\sigma_i) = \bm{X}_i^{\top} \bm{\beta} + Z_{i1}, \quad\text{ and }\quad
  \log(\kappa_i - x_0) = \bm{X}_i^{\top} \bm{\alpha} + \gamma Z_{i1} + Z_{i2}'.
\end{equation}
The starting point~$x_0$ and lower barrier~$\nu$ are set to be fixed depending
on the real data and clinical trial standard, which are shared for all subjects. 
In general, the subject with higher volatility and smaller reflected barrier is 
associated with higher risk of hypoglycemic event.

\section{Bayesian Inference} \label{sec:BayesEst}

Suppose that the event times of hypoglycemia are observed for a group of
$n$~patients during their follow-up times. For $i = 1, \ldots, n$,
let $t_{ij}$, $j = 1, \cdots, n_i$ be the $j$th of the $n_i$ observed gap times
of subject~$i$, the last of which is usually censored; let $\delta_{ij} = 1$ if
$t_{ij}$ is an event and~$0$ otherwise. In addition to a $p$-dimensional
covariate vector $\bm{X}_i$, the observed data include
$\bm{t}_i = (t_{i1}, \cdots, t_{in_i})^{\top}$ and
$\bm{\delta}_i = (\delta_{i1}, \cdots, \delta_{in_i})^{\top}$,
$i = 1, \ldots, n$. The parameters to be estimated are
$\bm{\Omega} = (\bm{\alpha}^{\top}, \bm{\beta}^{\top}, \theta_1, \theta_2', \gamma)^{\top}$.

\subsection{Likelihood, prior, and posterior} \label{subsec:lklh}

Given the covariate vector~$\bm{X}_i$ and subject-level 
frailties~$\bm{z}_i = (z_{i1}, z_{i2}')$, the gap times of subject~$i$
are assumed to be conditionally independent. The conditional likelihood
contribution of subject~$i$ given the unobserved frailties is
\begin{equation*}
  L_{i} (\bm{\Omega} | \bm{t}_i, \bm{\delta}_i , \bm{X}_i, \bm{z}_i)
    = \prod_{j = 1}^{n_i}
      \left[f(t_{ij}|x_0, \nu, \kappa_i, \sigma_i)\right]^{\delta_{ij}}
      \left[1 - F(t_{ij}|x_0, \nu, \kappa_i, \sigma_i)\right]^{ (1 - \delta_{ij})},
\end{equation*}
where the density and distribution functions are given by~\eqref{eq:dens} 
and~\eqref{eq:dist}, respectively, and $\sigma_i$ and $\kappa_i$ are defined 
in~\eqref{eq:model}.

In practice, the recorded event times are usually discrete in the
unit of days. This can be accounted for by assuming that the $t_{ij}$'s are
interval-censored. The contribution to the likelihood of subject~$i$ is then 
rewritten as
\begin{equation} 
  \label{eq:intvcondlklksubj}
  L_{i} (\bm{\Omega} | \bm{t}_i, \bm{\delta}_i , \bm{X}_i, \bm{z}_i)
  = \prod_{j = 1}^{n_i}
    \left[ F^*(t_{ij}|x_0, \nu, \kappa_i, \sigma_i) \right]^{\delta_{ij}}
    \left[ 1 - F \left(t_{ij} + \frac{1}{2} \Big\vert x_0, \nu, \kappa_i, \sigma_i) 
      \right) \right]^{(1 - \delta_{ij})},
\end{equation}
where
\begin{equation*}
  F^*(t_{ij}|x_0, \nu, \kappa_i, \sigma_i)  =
    F \left(t_{ij} + \frac{1}{2} \Big\vert x_0, \nu, \kappa_i, \sigma_i \right) -
    F\left(t_{ij} - \frac{1}{2} \Big\vert x_0, \nu, \kappa_i, \sigma_i \right).
\end{equation*}

Prior distributions of the parameters need to be specified to complete the
Bayesian model. For regression coefficients $\bm{\alpha}$ and $\bm{\beta}$, 
normal priors with zero mean and a large variance are specified. 
Inverse gamma priors are specified for the normal variances of the frailties
$\theta_1$ and $\theta_2'$. For the reparametrized coefficient of the frailty
$\gamma$, a vague normal prior with mean zero is imposed.
The priors are summarized as follows:
\begin{align}
  \label{eq:priors}
  \alpha_l &\sim N(0, \sigma^2_{\alpha}), \quad  l = 1, \ldots, p, \notag\\
  \beta_l &\sim N(0, \sigma^2_{\beta}), \quad l = 1, \ldots, p, \notag\\
  \gamma &\sim N(0, \sigma^2_{\gamma}),  \\
  \theta_1 &\sim \mbox{IG}(a, b), \notag\\
  \theta'_2 &\sim \mbox{IG}(a, b), \notag
\end{align}
where $N(0, \nu^2)$ is the normal distribution with mean zero and variance
$\nu^2 > 0$, and $\mbox{IG}(a, b)$ is the inverse gamma distribution with
shape~$a > 0$ and scale $b > 0$.
In our numerical studies, the hyper-parameters were set to be 
$\sigma^2_{\alpha} = \sigma^2_{\beta} = \sigma^2_{\gamma} = 10^2$, and 
$a = b = 1$.

With $\bm{z} = (\bm{z}_1, \cdots, \bm{z}_n)^{\top}$, 
combining the likelihood function and prior distributions leads to the
joint posterior density
\begin{align*}
  \pi( \bm{\Omega}, \bm{z}| \bm{t}_i, \bm{\delta}_i , \bm{X}_i) 
    & \propto
      \left[ \prod_{i=1}^{n} 
        L_{i} ( \bm{\Omega} | \bm{t}_i, \bm{\delta}_i , \bm{X}_i, \bm{z}_i ) 
        g(\bm{z}_i|\theta_1, \theta_2') \right]  \left[ \prod_{l = 1}^p q(\alpha_l) q(\beta_l) \right] 
      q(\theta_1) q(\theta_2') q(\gamma),
\end{align*}
where $q(\cdot)$ denotes a generic density function of its argument,
$L_{i} ( \bm{\Omega} | \bm{t}_i, \bm{\delta}_i , \bm{X}_i, \bm{z}_i )$ 
is the conditional likelihood function for subject~$i$ 
given in~\eqref{eq:intvcondlklksubj}, $g(\bm{z}_i|\theta_1, \theta_2')$ 
is the bivariate independent normal density of the reparametrized frailties of
subject~$i$, and all the $q(\cdot)$'s are priors given in~\eqref{eq:priors}.
Since all the priors are proper, the posterior is proper.

To make inferences about the parameters, we use MCMC. Draws from the posterior
distribution are drawn by a Gibbs sampling algorithm. The full conditional
distributions of all the parameters are sampled with the Metropolis--Hasting
algorithm because the FHT density is not of any existing standard form. We
incorporated our customized FHT distribution into a generic implementation
through \proglang{R} package~\pkg{NIMBLE} \citep{devalpine2017programming}.
Due to the large number of unobserved frailties, the resulting chains are highly
auto-correlated. A good number of draws can be obtained after thinning those
long chains.

\subsection{Model selection}
With Dev$(\bm{\Omega})$ denoting the deviance calculated by negated
observed-data
loglikelihood evaluated at $\bm{\Omega}$, DIC is given as
\begin{equation*}
  \mbox{DIC} = \mbox{Dev}( \bar{\bm{\Omega}} ) + 2p_D,
\end{equation*}
where $\bar{\bm{\Omega}}$ is the posterior mean of $\bm{\Omega}$,
$p_D = \overline{\mbox{Dev}}(\bm{\Omega}) - \mbox{Dev}(\bar{\bm{\Omega}})$ is 
the effective number of parameters, and  $\overline{\mbox{Dev}}(\bm{\Omega})$ is
the posterior mean of $\mbox{Dev}(\bm{\Omega})$.
This calculation based on the observed-data likelihood is the
$\mbox{DIC}_3$ in \citet{celeux2006deviance}. Since there is
no closed-form of the observed-data likelihood, we used
Monte Carlo integration to approximate it.
The observed-data likelihood of subject~$i$ has the form 
\begin{equation*}
    \label{eq:obsdatlklksubj}
    L_{i} (\bm{\Omega} | \bm{t}_i, \bm{\delta}_i , \bm{X}_i) 
    = \int L_{i} (\bm{\Omega} | \bm{t}_i, \bm{\delta}_i , \bm{X}_i, \bm{z}_i)
      g(\bm{z}_i|\theta_1, \theta_2') \dd \bm{z}_i,  
\end{equation*}
where $L_{i} (\bm{\Omega} | \bm{t}_i, \bm{\delta}_i , \bm{X}_i, \bm{z}_i)$ is given 
in~\eqref{eq:intvcondlklksubj}.
The Monte Carlo approximation for the observed-data likelihood for subject~$i$
is
\begin{align}
  \label{eq:MCapprox}
  L_{i} (\bm{\Omega} | \bm{t}_i, \bm{\delta}_i , \bm{X}_i) 
    \approx \frac{1}{M} \sum_{m=1}^M 
      L_{i} (\bm{\Omega} | \bm{t}_i, \bm{\delta}_i, \bm{X}_i, \bm{z}_i^{(m)} ),
\end{align}
where $\bm{z}_i^{(1)}, \cdots, \bm{z}_i^{(M)}$ are Monte Carlo samples 
that each can be easily generated by the independent normal distributions, and
$M$ is the Monte Carlo sample size. The deviance is given by 
\begin{equation*}
  \mbox{Dev} (\bm{\Omega}) = -2 \sum_{i=1}^n \log 
  L_{i} (\bm{\Omega}|\bm{t}_i, \bm{\delta}_i, \bm{X}_i).
\end{equation*}
Models with smaller DIC values are preferred.

The other criterion LPML is calculated based on conditional 
predictive ordinate (CPO). Let 
$D_{-i} = \{(\bm{t}_j, \bm{\delta}_j, X_j): j = 1,\ldots, n; j \ne i \}$,
denote the observed data with the $i$th~subject excluded. The CPO for the
$i$th~subject is the leave-one-out predictive likelihood
\begin{align*}
  \mbox{CPO}_i = 
    \int L_{i} (\bm{\Omega}|\bm{t}_i, \bm{\delta}_i, \bm{X}_i) q(\bm{\Omega}|D_{-i}) \dd \bm{\Omega},
\end{align*}
where
$q(\bm{\Omega}|D_{-i})$ is the marginal posterior distribution of $\bm{\Omega}$ 
with the $i$th subject excluded. The Monte Carlo estimate of
$\mbox{CPO}_i$ \citep{dey1997bayesian} is
\begin{equation}
  \label{eq:cpo}
   \widehat{\mbox{CPO}}_{i} = \left[\frac{1}{K}\sum_{k=1}^K 
   \frac{1}{L_{i} (\bm{\Omega}|\bm{t}_i, \bm{\delta}_i, \bm{X}_i)}\right]^{-1},
\end{equation}
where
$\bm{\Omega}^{(k)} = (\bm{\alpha}^{(k)}, \bm{\beta}^{(k)}, \theta_{1}^{(k)}, \theta_{2}'^{(k)}, \gamma^{(k)})^{\top}$, 
$k = 1, \cdots, K$, is the $k$th iteration of posterior draws.
Each term $L_i$ in Equation~\eqref{eq:cpo} can be approximated the same way as
in~\eqref{eq:MCapprox}. Then the LPML can be calculated by
\begin{equation*}
   \widehat{\mbox{LPML}} = \sum_{i = 1}^n \log (\widehat{\mbox{CPO}}_i).
\end{equation*}
Models with higher LPML values are preferred.

\section{Simulation} \label{sec:simulation}

Simulation studies were conducted to evaluate the performance of the Bayesian
estimator and the model selection criteria. To mimic the real application
analyzed in the next section, we considered a setting of two covariates for each
subject~$i$: $X_{i1}$, baseline insulin level and $X_{i2}$, baseline body mass
index (BMI). A bivariate gamma distribution with a normal copula was fitted to
the real data with \proglang{R} package \pkg{copula}
\citep{hofert2018elements}. The covariates were then generated from the fitted
bivariate gamma distribution.

Three frailty models were used to generate recurrent events with covariate
vector $\bm{X}_i = (1, X_{i1}, X_{i2})^{\top}$.
Model~1 is the full model~\eqref{eq:model} with
$\bm{\alpha} = (2.9, 0.2, -0.1)^{\top}$,
$\bm{\beta} = (0.9, -0.2, -0.1)^{\top}$, and $\gamma = -0.55$;
the frailties $z_{i1}$ and $z_{i2}'$ were generated from 
the normal distribution with mean~zero and variance $\theta_1 = 0.2$ 
and $\theta_2' = 0.3$, respectively. The starting point~$x_0$ and
the lower barrier~$\nu$ were set to be $x_0 = 10$ and 
$\nu = 3.9$, respectively, which will be discussed in Section~\ref{sec:aplc}. 
Model~2 is the same as Model~1 except that $\gamma = 0$, which means that the
frailties in the upper reflection barrier and the volatility are independent.
Model~3 is also a reduced model of Model~1 with the restrictions
$\gamma = -1$ and $\theta_2' = 0$, which implies that the 
upper reflection barrier and the volatility share the same frailty. 
The negative sign of gamma was chosen as suggested by the real data analysis.
For ease of reference, the
three models are referred to as, respectively, correlated-frailty,
independent-frailty, and shared-frailty.

For each model,
the gap times were generated with the rejection sampling algorithm
in Section~\ref{subsec:FHT} until the sum of generated gap times
is equal to or larger than the follow-up time of each subject~$i$, where the
follow-up times were independently generated from the empirical distribution of
the follow-up times of the real data.
The generated gap times were then rounded into integers as days. 
Two levels of sample sizes, $n \in \{200, 400\}$ were considered. 
For each simulation setting, $200$~datasets were generated.

The Bayesian analysis of each dataset used prior distributions specified in
Equation~\eqref{eq:priors}. For the coefficient parameters, we set
$\sigma^2_{\alpha} = \sigma^2_{\beta} = \sigma^2_{\gamma} = 10^2$, which leads
to vague but proper prior distributions. The hyper-parameters $(a, b)$ of the
inverse gamma prior for the frailty variances are often set to  be $a = b = 1$
\citep{gelman2006prior, manda2005bayesian}. Given a dataset, an MCMC was run
with $90,000$ iterations. The first $15,000$ iterations were discarded as
burn-in period and the rest were thinned by $150$. The convergence of MCMC was
checked with Heidelberger \& Welch's diagnostic
\citep{heidelberger1983simulation}, which is available in \proglang{R}
package~\pkg{CODA} \citep{plummer2006coda}. The posterior inferences and model
selection criteria were obtained based on the resulting $500$ MCMC samples.
The choice of posterior sample size, $500$, was driven by the consideration to
reduce computational cost for the calculation of DIC/LPML criteria. Those
criteria were calculated based on observed likelihood that involves integrating
over the random effect with Monte Carlo approximation, which is time-consuming.
To further justify the independence of the posterior samples, the effective
sample size was calculated for the $500$ MCMC samples using the \proglang{R}
package~\pkg{CODA}. The average effective sample size is at least $332$ among
parameters of different models across $200$ simulation replicates, which we
believe is reasonable for inferences.

\begin{table}[tbp]
  \caption{Results of parameter estimation under correct specifications
    for three models with sample size $n \in \{200, 400\}$. 
    SD, posterior standard deviation; ESD, empirical standard deviation; 
    CR, coverage rate of $95\%$ HPD credible intervals.}  
  \label{tab:simuResSum}
  \begin{center}
  \begin{tabular}{crrrrrrrrrrr}
    \toprule
            &  &  & \multicolumn{4}{c}{$n = 200$} & \multicolumn{4}{c}{$n = 400$} \\
                  \cmidrule(lr){4-7} \cmidrule(lr){8-11}
      Model & Para  & True & Bias & SD & ESD & CR & Bias & SD & ESD & CR \\ 
  \midrule
  \multirowcell{2}{correlated-\\frailty} 
  &$\alpha_0$ & 2.90 & 0.090 & 0.166 & 0.171 & 0.90 & 0.050 & 0.105 & 0.112 & 0.92 \\ 
  &$\alpha_1$ & 0.20 & 0.051 & 0.151 & 0.152 & 0.92 & 0.019 & 0.096 & 0.097 & 0.96 \\ 
  &$\alpha_2$ & $-$0.10 & 0.004 & 0.111 & 0.104 & 0.96 & $-$0.001 & 0.072 & 0.073 & 0.96 \\ 
  &$\gamma$ & $-$0.55 & 0.011 & 0.298 & 0.257 & 0.96 & $-$0.019 & 0.199 & 0.203 & 0.96 \\ 
  &$\beta_0$ & 0.90 & $-$0.016 & 0.047 & 0.048 & 0.92 & $-$0.003 & 0.032 & 0.033 & 0.95 \\ 
  &$\beta_1$ & $-$0.20 & $-$0.006 & 0.052 & 0.048 & 0.96 & 0.002 & 0.035 & 0.032 & 0.96 \\ 
  &$\beta_2$ & $-$0.10 & 0.001 & 0.048 & 0.051 & 0.93 & $-$0.001 & 0.033 & 0.033 & 0.95 \\ 
  &$\theta_1$ & 0.20 & 0.029 & 0.038 & 0.036 & 0.90 & 0.010 & 0.025 & 0.026 & 0.92 \\ 
  &$\theta_2'$ & 0.30 & 0.129 & 0.140 & 0.115 & 0.89 & 0.057 & 0.086 & 0.077 & 0.92 \\ 
  [1ex]
  \multirowcell{2}{independent-\\frailty} 
  & $\alpha_0$ & 2.90 & 0.076 & 0.128 & 0.110 & 0.95 & 0.037 & 0.084 & 0.078 & 0.94 \\ 
  & $\alpha_1$ & 0.20 & 0.052 & 0.149 & 0.143 & 0.95 & 0.017 & 0.099 & 0.096 & 0.94 \\ 
  & $\alpha_2$ & $-$0.10 & $-$0.005 & 0.107 & 0.096 & 0.96 & $-$0.002 & 0.071 & 0.072 & 0.92 \\ 
  & $\beta_0$ & 0.90 & $-$0.010 & 0.047 & 0.045 & 0.96 & $-$0.008 & 0.033 & 0.031 & 0.94 \\ 
  & $\beta_1$ & $-$0.20 & $-$0.009 & 0.053 & 0.050 & 0.93 & 0.000 & 0.036 & 0.038 & 0.96 \\ 
  & $\beta_2$ & $-$0.10 & 0.006 & 0.048 & 0.049 & 0.94 & $-$0.001 & 0.033 & 0.034 & 0.93 \\ 
  & $\theta_1$ & 0.20 & 0.029 & 0.037 & 0.033 & 0.92 & 0.015 & 0.025 & 0.023 & 0.92 \\ 
  & $\theta_2'$ & 0.30 & 0.102 & 0.126 & 0.099 & 0.95 & 0.062 & 0.083 & 0.077 & 0.90 \\ 
  [1ex]
  \multirowcell{2}{shared-\\frailty}  
  & $\alpha_0$ & 2.90 & 0.042 & 0.126 & 0.124 & 0.97 & 0.011 & 0.082 & 0.081 & 0.95 \\ 
  & $\alpha_1$ & 0.20 & 0.022 & 0.104 & 0.105 & 0.93 & 0.005 & 0.069 & 0.072 & 0.94 \\ 
  & $\alpha_2$ & $-$0.10 & $-$0.002 & 0.078 & 0.065 & 0.98 & 0.001 & 0.052 & 0.051 & 0.94 \\ 
  & $\gamma$ & $-$1.00 & $-$0.038 & 0.202 & 0.188 & 0.98 & 0.011 & 0.134 & 0.143 & 0.93 \\ 
  & $\beta_0$ & 0.90 & $-$0.018 & 0.046 & 0.051 & 0.90 & $-$0.012 & 0.032 & 0.031 & 0.94 \\ 
  & $\beta_1$ & $-$0.20 & $-$0.008 & 0.050 & 0.053 & 0.92 & $-$0.003 & 0.035 & 0.036 & 0.93 \\ 
  & $\beta_2$ & $-$0.10 & $-$0.003 & 0.046 & 0.044 & 0.95 & $-$0.004 & 0.032 & 0.032 & 0.92 \\ 
  & $\theta_1$ & 0.20 & 0.025 & 0.036 & 0.034 & 0.90 & 0.013 & 0.024 & 0.020 & 0.96 \\ 
  \bottomrule
  \end{tabular}
  \end{center}
\end{table}

Table~\ref{tab:simuResSum} summarizes the results of the Bayesian
estimation for all three models when fitted with correct specifications. The
posterior mean of each parameter was considered as the point estimator.
The empirical bias of the estimates for all parameters is close to zero except
that of $\theta_2$ when $n = 200$ for the correlated-frailty and the
independent-frailty model; in both cases the bias reduces as $n$ increases.
The mean of the posterior standard deviation of the estimates agrees
closely with the empirical standard deviation of the point
estimates for most parameters even for sample size $n = 200$. Consequently, the
empirical coverage rates of the $95\%$ highest posterior density (HPD)
credible intervals are close to the nominal level, and the agreement improves in
general as $n$ increases from 200 to 400.

\begin{table} [tbp]
  \caption{Model comparison result with DIC and LPML with sample size 
  $n \in \{200, 400, 800, \mbox{and}, 1600\}$.
  Freq (\%): frequency of the correct model being selected; 
  Mean: average of the DIC or LPML.}
  \label{tab:simuModCprSum}
  \begin{center}
  \begin{tabular}{cccrrrrrr}
  \toprule
  &   &  &
  \multicolumn{2}{c}{correlated-frailty}  &
  \multicolumn{2}{c}{independent-frailty}  &
  \multicolumn{2}{c}{shared-frailty} \\
  \cmidrule(lr){4-5} \cmidrule(lr){6-7} \cmidrule(lr){8-9} 
  True Model  & Criterion & $n$ & Freq & Mean & Freq & Mean & Freq & Mean \\ 
  \midrule
  correlated- & DIC   & 200  & 61 &  11889.5   & 23 &  11898.0 & 16 & 11900.9 \\ 
  frailty     &       & 400  & 81 &  23812.5   & 14 &  23829.4 & 5 &  23844.9 \\
              &       & 800  & 94 &  47324.2   &  4 &  47364.0 & 2 &  47394.8 \\
              &       & 1600 & 95 &  94528.6   &  1 &  94612.9 & 4 &  94741.0 \\ 
        & LPML  & 200  & 50 & $-$5957.2  & 27 & $-$5960.7  & 23 & $-$5961.7  \\ 
        &       & 400  & 68 & $-$11922.6 & 19 & $-$11931.2 & 13 & $-$11936.6 \\ 
        &       & 800  & 87 & $-$23682.2 &  7 & $-$23702.5 &  6 & $-$23715.7 \\ 
        &       & 1600 & 90 & $-$47292.0 &  4 & $-$47332.0 &  6 & $-$47395.2 \\  
  [1ex]
  independent- & DIC & 200  & 22 &  10885.3 & 71 &  10884.2 & 7 &  10901.7 \\   
  frailty      &     & 400  & 23 &  21701.9 & 75 &  21700.4 & 2 &  21747.0 \\ 
               &     & 800  & 32 &  43296.2 & 68 &  43295.4 & 0 &  43391.4 \\ 
               &     & 1600 & 28 &  86551.1 & 72 &  86549.6 & 0 &  86744.6 \\ 
          & LPML & 200  & 19 & $-$5448.0  & 72 & $-$5446.6  & 9 & $-$5457.0 \\ 
          &      & 400  & 18 & $-$10857.3 & 79 & $-$10855.7 & 3 & $-$10880.8 \\
          &      & 800  & 25 & $-$21655.6 & 74 & $-$21653.6 & 1 & $-$21703.5 \\ 
          &      & 1600 & 19 & $-$43284.7 & 81 & $-$43282.6 & 0 & $-$43381.7 \\
  [1ex]
  shared-   & DIC & 200  & 7  & 12415.9 & 0 & 12445.3 & 93 & 12405.9 \\ 
  frailty   &     & 400  & 9  & 24773.1 & 0 & 24844.2 & 91 & 24758.6 \\
            &     & 800  & 11 & 49302.4 & 0 & 49453.9 & 89 & 49282.7 \\
            &     & 1600 & 23 & 98722.2 & 0 & 99038.5 & 77 & 98702.1 \\ 
          & LPML & 200  & 11 & $-$6221.9  & 2 & $-$6238.7  & 87 & $-$6214.7 \\
          &       & 400  & 16 & $-$12410.4 & 1 & $-$12448.2 & 83 & $-$12401.2 \\
          &       & 800  & 22 & $-$24680.6 & 1 & $-$24760.0 & 77 & $-$24671.1 \\
          &       & 1600 & 30 & $-$49397.1 & 0 & $-$49556.2 & 70 & $-$49386.0 \\ 
  \bottomrule
  \end{tabular} 
  \end{center}
  \end{table}

To evaluate the performance of model comparison criteria, we fitted three models
to each dataset. DIC and LPML were used to select among the three fitted models.
The Monte Carlo sample size used for approximating the integrals was set to be
$M = 500$. Table~\ref{tab:simuModCprSum} summarizes the frequencies under each
criterion that the correct model was selected based on either DIC or LPML for
sample size $n \in \{200, 400, 800, \mbox{and}, 1600\}$. On average, the
correctly specified models have the smallest DIC and highest LPML. The results
suggest good performance of both DIC and LPML in selecting the right models. For
example, when correlated-frailty model is the data generating model, the
proportion of correctly identify the true model increases for both DIC and LPML
as the sample size increasing, which are $95\%$ and $90\%$, respectively, 
with DIC slightly outperforming LPML. When the two reduced models are the data
generating models, both DIC and LPML still select them with the highest
frequency, but the tendency to choose the full model also increases when the
sample size increases. Such observations echo the limitations of DIC and LPML in
distinguishing the true model and an overfitted model \citep{maity2021bayesian}.
In our application, both criteria select either the correct model or the full
model, which still provides valuable information for practitioners.


\section{Hypoglycemic Event Time Analysis} \label{sec:aplc}

\begin{table}[tbp]
  \caption{Summary of the covariates from the DURABLE trial. BMI, body mass index; 
  BP, blood pressure; SD, standard deviation.}
  \label{tab:datasum}
  \begin{center}
  \begin{tabular}{lrrrrr} 
    \toprule
        Variable                        & Minimum & Median & Maximum  & Mean  & SD\\ 
    \midrule
      Fasting glucose (mmol/l)          & 0.23 & 10.45 & 25.96 & 10.78 & 3.72 \\
      Adiponectin (\si{\micro\gram}/mL) & 0.01 & 5.57 & 49.01 & 6.99 & 5.52 \\
      Fasting insulin (mIU/L)           & 0.18 & 7.91 & 142.68 & 10.40 & 9.81 \\
      Height (cm)                       & 124.25 & 166.44 & 198.09 & 166.47 & 10.71 \\
      BMI (kg/$\mbox{m}^2$)             & 15.88 & 31.28 & 62.62 & 31.71 & 6.18 \\ 
      Diastolic BP (mmHg)               & 45.01 & 78.70 & 116.30 & 78.23 & 9.46 \\
      Systolic BP (mmHg)                & 47.26 & 130.02 & 196.67 & 131.53 & 16.11 \\
      Heart rate  (beats per minute)    & 43.86 & 76.61 & 121.05 & 76.76 & 9.82 \\
      Duration diabetes (years)         & 0.03 & 8.57 & 39.48 & 9.75 & 6.17 \\
    \bottomrule
  \end{tabular}
\end{center}
\end{table}

The proposed model was applied to analyze the hypoglycemic event times from the
DURABLE trial \citep{buse2009durability}. Between $2005$ and $2007$, 
$2187$ patients with type~2 diabetes from $11$ countries were enrolled 
in the study. The dataset contains the possibly censored times of hypoglycemic
events of the patients during their follow-up periods. Also, available are a
collection of baseline covariates, which allows assessments of risk factors of
hypoglycemia among the patients. The median follow-up time of the 
patients is $168$ days. Continuous baseline covariates include
fasting blood glucose, fasting insulin, adiponectin, weight, height, 
body mass index (BMI), systolic blood pressure, diastolic blood pressure, 
heart rate, and duration of diabetes. Summaries of the continuous covariates 
are presented in Table~\ref{tab:datasum}. Three important variables, fasting
glucose level, adiponectin level, and fasting insulin level, have extremely
high values, which calls for prudence in data analysis. Two categorical
variable are available. The first one is starter insulin regimens with two
levels, twice-daily lispro mix $75/25$ (LM$75/25$; $75\%$ lispro protamine
suspension, $25\%$ lispro) versus once-daily insulin glargine. The second
one is the usage of oral antihyperglycemic drugs with three levels,
thiazolidinedione, sulfonylurea, and both. All the available covariates are 
subject-level and time-independent covariates.

After excluding the subjects with missingness in covariates or outside reference
range, the dataset contains $n = 1943$ patients. Prior to model fitting, all the
continuous covariates were standardized. Log transformation was applied to two
right-skewed covariates, baseline adiponectin and baseline fasting insulin,
before standardization. Among the $1943$ patients, $570$ ($29$\%) received both
oral antihyperglycemic drugs, $1207$ ($62$\%)  only received sulfonylurea, and
$166$ ($9$\%) only received thiazolidinedione. For ease of discussion, the group
that received both of two drugs were used as the reference group; two dummy
variables, \textsf{sulf-Only}, which is $1$ if only received sulfonylurea, and
\textsf{tzd-only}, which is $1$ if only received thiazolidinedione, were
included. Define an indicator variable for the insulin regime \textsf{LM}, which
equals~1 for the $959$ ($49$\%) patients who received LM $75/25$ and~0 for the
$984$ ($51$\%) patients who received glargine. Some patients had multiple
hypoglycemic events within a single calendar date. In this case, the gap times
between successive hypoglycemic events were recorded as zero. This is handled by
treating the gap times in days as interval-censored with the likelihood
constructed with~\eqref{eq:intvcondlklksubj} in Section~\ref{subsec:lklh}. The
daily hypoglycemic event rates of the patients have a wide range from $0$ to
$0.77$ with mean $0.07$. These descriptive statistics indicate the existence of
severe heterogeneity risk of hypoglycemia among subjects.

\begin{table} 
  \caption{Model comparison for three models (correlated-frailty,
    independent-frailty, and shared-frailty model) fitted to the DURABLE data.}
  \label{tab:aplcModCpr}
  \begin{center}
  \begin{tabular}{lrrr}
    \toprule
            &  correlated-frailty & independent-frailty  
            &  shared-frailty \\
    \midrule
      DIC  & 131400.8   &  131397.0  & 131944.6   \\
      LPML & $-$65707.5 & $-$65706.7 & $-$65986.7 \\
    \bottomrule
  \end{tabular}
  \end{center}
\end{table}

The three models along with their priors investigated in
Section~\ref{sec:simulation} were fitted to the DURABLE data. The lower boundary
was set to $3.9$ mmol/l ($70$ mg/dl), which is the clinical standard for
hypoglycemic events \citep{seaquist2013hypoglycemia}. The starting point~$x_0$
of the Brownian motion after each hypoglycemic event was set to be $10$, which
is the rounded integer of the mean of the baseline fasting glucose level of all
the patients. The sensitivity analysis of different priors and starting
points have been conducted and the results showed that the estimates of the 
covariate effects are stable, which can be found in the Supplementary
Materials. For each model, an MCMC was run for $55,000$ iterations and thinned
by $10$ after discarding the first $15,000$ iterations as burn-in. The
convergence of the MCMC chains was monitored by trace plots. The results of DIC
and LPML for the three models are presented in Table~\ref{tab:aplcModCpr}. Both
criteria suggest that the correlated-frailty model and the independent-frailty
model are similar, both of which are preferred to shared-frailty model. Given
that the correlated-frailty and the independent-frailty have close model fit, we
chose the independent-frailty model as it is parsimonious. That is, the two
frailties in the upper reflection barrier and the volatility could be treated as
independent.

\begin{sidewaystable} [tbp]
\centering
\caption{Estimated parameters of the independent-frailty model. 
BMI, body mass index; BP, blood pressure;
SD, standard deviation; 
CI, $95$\% HPD credible interval or $95$\% confident interval.}
\label{tab:aplcres}
\begin{tabular}{lrrcrrcrrc}
    \toprule
        & \multicolumn{6}{c}{Independent-frailty Model} 
        & \multicolumn{3}{c}{Proportional hazards gap time} \\
        \cmidrule(lr){2-7}   \cmidrule(lr){8-10} 
        & \multicolumn{3}{c}{Volatility}  
        & \multicolumn{3}{c}{Upper reflection barrier} \\
                    \cmidrule(lr){2-4}   \cmidrule(lr){5-7} 
    Covariates  & Mean & SD & 95\% CI 
                & Mean & SD & 95\% CI 
                & Mean & SD & 95\% CI \\ 
    \midrule                 
    Intercept         & 0.910    & 0.036 & \textbf{[0.840, 0.982]} 
                      & 2.868    & 0.067 & \textbf{[2.745, 2.999]} 
                      & - & - & -  \\ 
    Fasting glucose   & $-$0.054 & 0.019 & \textbf{[$-$0.096, $-$0.017]} 
                      & 0.108    & 0.031 & \textbf{[0.049, 0.170]} 
                      & $-$0.122 & 0.024 & \textbf{[$-$0.169, $-$0.075]}\\ 
    Adiponectin       & 0.046    & 0.019 & \textbf{[0.011, 0.085]} 
                      & 0.030    & 0.031 & [$-$0.035, 0.088]       
                      & 0.042    & 0.026 & [$-$0.008, 0.093] \\ 
    Fasting insulin   & $-$0.107 & 0.022 & \textbf{[$-$0.150, $-$0.065]} 
                      & 0.169    & 0.032 & \textbf{[0.108, 0.231]} 
                      & $-$0.208 & 0.026 & \textbf{[$-$0.259, $-$0.157]} \\  
    Height            & $-$0.047 & 0.019 & \textbf{[$-$0.085, $-$0.011]} 
                      & 0.004    & 0.032 & [$-$0.056, 0.068]       
                      & $-$0.059 & 0.024 & \textbf{[$-$0.107, $-$0.012]} \\ 
    BMI               & $-$0.089 & 0.021 & \textbf{[$-$0.132, $-$0.046]} 
                      & $-$0.113 & 0.033 & \textbf{[$-$0.180, $-$0.050]} 
                      & $-$0.058 & 0.026 & \textbf{[$-$0.109, $-$0.006]} \\
    Diastolic BP      & $-$0.069 & 0.022 & \textbf{[$-$0.110, $-$0.025]} 
                      & 0.055    & 0.036 & [$-$0.018, 0.122] 
                      & $-$0.109 & 0.028 & \textbf{[$-$0.164, $-$0.053]} \\  
    Systolic BP       & 0.025    & 0.020 & [$-$0.012, 0.064] 
                      & $-$0.027 & 0.032 & [$-$0.094, 0.033] 
                      & 0.041 & 0.027 & [$-$0.013, 0.094] \\ 
    Heart rate        & 0.008    & 0.018 & [$-$0.027, 0.042] 
                      & 0.048    & 0.029 & [$-$0.009, 0.104] 
                      & $-$0.011 & 0.024 & [$-$0.058, 0.036] \\
    Duration diabetes & 0.081    & 0.018 & \textbf{[0.049, 0.117]} 
                      & $-$0.049 & 0.028 & [$-$0.105, 0.004] 
                      & 0.120 & 0.023 & \textbf{[0.074, 0.166]} \\ 
    LM                & 0.168    & 0.036 & \textbf{[0.098, 0.240]} 
                      & $-$0.102 & 0.061 & [$-$0.219, 0.017] 
                      & 0.236 & 0.046 & \textbf{[0.145, 0.326]} \\ 
    tzd-only          & $-$0.490 & 0.072 & \textbf{[$-$0.628, $-$0.351]} 
                      & 0.322    & 0.149 & \textbf{[0.035, 0.614]} 
                      & $-$0.673 & 0.099 & \textbf{[$-$0.868, $-$0.479]} \\
    sulf-only         & $-$0.022    & 0.037 & [$-$0.093, 0.054] 
                      & 0.059    & 0.066 & [$-$0.067, 0.195] 
                      & $-$0.064 & 0.057 & [$-$0.175, 0.047] \\
                       [1em]
    Frailty Variance  & 0.408    & 0.026 & \textbf{[0.356, 0.458]} 
                      & 0.535    & 0.044 & \textbf{[0.450, 0.620]} 
                      & 0.759 & - & - \\ 
    \bottomrule
\end{tabular}
\end{sidewaystable}

Table~\ref{tab:aplcres} summarizes the estimated model parameters, their
standard errors, and 95\% HPD credible confidence intervals from the fitted
independent-frailty model. The results from the volatility model suggest that
patients with higher baseline fasting blood glucose level, lower adiponectin,
higher fasting insulin, higher height, higher BMI, higher diastolic BP, lower
duration of diabetes are significantly associated with lower volatility and,
hence, lower risk of hypoglycemia. Patients who received LM $75/25$ appear to
have higher volatility or higher risk of hypoglycemia compared to those who
received glargine. For the oral antihyperglycemic drugs, patients who received
only thiazolidinedione appear to have lower volatility or lower risk of
hypoglycemia compared to those who received both thiazolidinedione and
sulfonylurea; patients who received only sulfonylurea are not significantly
different from those who received both.

\begin{figure} [tbp]
\centering
\includegraphics[max width=\textwidth]{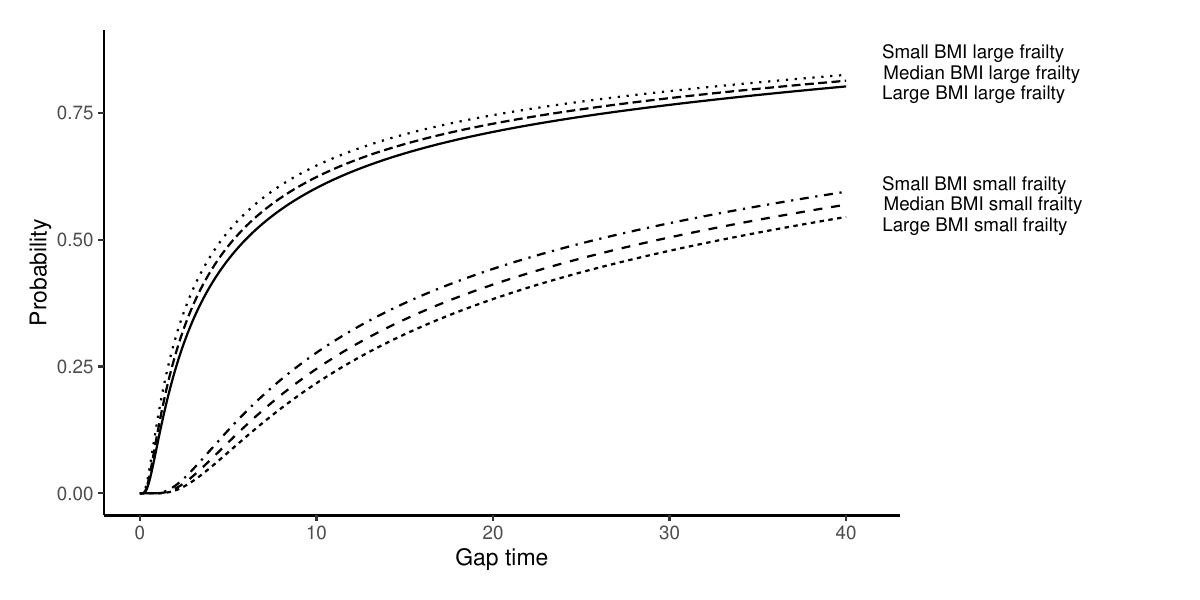}  
\caption{First hitting time distribution with fitted parameters of
independent-frailty model. Distribution functions are derived from $6$ 
combinations of $3$ levels of BMI, small, median, and large, which represent the
$25\%$, $50\%$, and $75\%$ quantile of the standardized BMI, respectively; and
$2$ levels of frailties, small and large, which represent the $25\%$ and $75\%$
quantile of the frailty distributions both in volatility $\sigma$ and upper
reflection barrier $\kappa$. Other covariates remain the same at their median
level after being standardized.}
\label{fig:predictivePlotsBMI}
\end{figure}

In the upper reflection barrier model, fewer covariates are significant and they
are a subset of those that are significant in the volatility model. Patients
with higher baseline fasting blood glucose level, higher fasting insulin, and
lower BMI are associated higher reflection barrier and, hence, lower risk of
hypoglycemia. For the oral antihyperglycemic drugs, patients who received only
thiazolidinedione appear to have higher reflection barrier and, hence, lower
risk of hypoglycemia compared to those who received both thiazolidinedione and
sulfonylurea. Interestingly, the effect of baseline fasting blood glucose level,
fasting insulin, and received only thiazolidinedione, the coefficients have the
same direction on the risk of hypoglycemia in the models for volatility and
upper reflection barrier (with opposite coefficient signs). In contrast, BMI is
significant in affecting both the volatility and the upper reflecting barrier,
but with opposite directions (with the same coefficient signs). That is, the
overall effect of BMI on the risk of hypoglycemia is complicated by lowering the
volatility (or lowering the risk) while decreasing the upper reflection barrier
(or increasing the risk). This discovery has not been reported in the quantile
regression analysis of \citet{ma2021heterogeneous}. But the overall effects of 
baseline BMI is worth further investigating.
Figure~\ref{fig:predictivePlotsBMI} gives a set of the first hitting time
distributions with the fitted parameters of independent-frailty model using 
different levels of baseline BMI and frailties. 
From Figure~\ref{fig:predictivePlotsBMI}, regardless of the frailty
level, smaller BMI is associated with a higher risk of having hypoglycemic
events.

Comparison of the estimates between the proposed model and proportional hazards
model of gap time between recurrent events is also given in
Table~\ref{tab:aplcres}. With the exception of the covariate Adiponectin, the
other covariates exhibit similar levels of significance between proportional
hazards model and volatility in the proposed model. For Adiponectin, the
estimates in the volatility and upper reflection barrier model are $0.046
[0.011, 0.085]$ and $0.030 [-0.035, 0.088]$, respectively, in the proposed
model. This suggests that Adiponectin contributes in different directions on
volatility and upper reflection barrier model for the recurrent risk. Given that
the volatility component exerts a stronger influence on recurrence, our estimate
indicates that the higher value of Adiponectin is associated with an increased
risk of recurrence. In proportional hazards model, the estimate of Adiponectin
is $0.042 [-0.008, 0.093]$, aligning with the trend observed in the proposed
model. For the other significant covariates, the signs of estimates between the
volatility model and proportional hazards model are consistent. This observation
suggests that, for this dataset, the proposed model offers a more detailed
characterization of the concealed glucose levels linked to recurrent events.

The two independent frailties in the volatility and the upper reflection barrier
capture much of the heterogeneity among the patients beyond the covariates. The
variances of the two frailties are estimated to be far away from zero. From the
fitted model, the range of volatility spans from $0.65$ to $9.52$ with
median~$2.75$ and mean~$3.03$; the range of upper reflection barrier spans from
$12.69$ to $83.47$ with median~$29.57$ and mean $29.88$. The model without the
frailties fits much poorer in terms of DIC and LPML (not reported).

\section{Discussion} \label{sec:conclusion}
The risk of hypoglycemia is an important concern in diabetes management. It is
natural to model the underlying blood glucose level as hypoglycemia occurs when
it hits a lower boundary and hyperglycemia occurs when it hits an upper
boundary. Because of the unique setting where hyperglycemia cannot be reliably
observed in self-reported data, it is challenging to model the blood glucose
level as a stochastic process. The proposed Brownian motion model with an upper
reflection barrier allows bypassing the need for observing hyperglycemic event
times. Only hypoglycemic times are needed for the model fitting. This model
fitting is made possible by the FHT density and distribution
\citep{hu2012hitting}. The recurrence of the hypoglycemic events is captured by
a sequence of stochastic processes reaching the lower-boundary. The upper
reflection barrier and volatility of the reflected Brownian motion are linked to
patient-level covariates and frailties. Due to the unobserved frailties, we
resorted to Bayesian inference for the parameters with MCMC implemented with
\pkg{NIMBLE} \citep{devalpine2017programming}. The computation of our work
relies on an accurate implementation of the FHT density/distribution functions
as well as the rejection sampling algorithm with the 3-piece proposal kernel.
Another computation challenge is the complexity brought by unobserved frailties
in calculating model selection criteria DIC and LPML. We applied Monte Carlo
integration for an approximation, with which the two criteria were shown to be
reasonably effective  in selecting the correct frailty model by simulation
studies.

It is worthwhile to revisit the key model assumptions. In our model, the imposed
upper reflection barrier reflects the expectation that the blood sugar level is
bounded and will not reach infinity. This modeling choice aids in illustrating
that glucose levels will ultimately be controlled (dropping down) in the range.
We believe this assumption is reasonable, especially for participants who are
consistently under regular blood sugar maintenance. Another key assumption is
that the glucose level returns to normal immediately following the occurrence of
the hypoglycemia event. In reality, it certainly takes time to consume food and
bring back the glucose level, but the recovery is usually in minutes and, thus,
quick enough so we neglect the time used. We also assumed that, after a
hypoglycemia event, the glucose level restarts at a fixed level~$x_0$. A
sensitivity analysis with different values of~$x_0$ showed little difference in
the resulting regression coefficient estimates.

Under the model framework, a subject with larger volatility and lower upper
reflection barrier is associated with higher risk of the hypoglycemia event. In
our experience, the covariate with the same direction on the risk of
hypoglycemia (with opposite coefficient signs in volatility and upper reflection
barrier) is usually consistent with that of the classic gap time models.
On the other hand, if a covariate contributes a positive (negative) effect to
both the volatility and upper reflection barrier, its overall impact is less
straightforward. In this case, it is possible that the covariate shows no
significant impact from the classic gap time models, while play an import role
for volatility or upper reflection barrier. An example of this is the covariate
Adiponectin in our data application. Therefore, despite its complexity, the
proposed model provides additional insights for a better understanding of the
data. To further investigate the covariate overall effect of proposed model, we
recommend plotting the FHT distribution to provide an overall characterization
of the impact of this covariate.

Several directions are worth further investigation.
In the broad sense, the proposed model can be applied
to scenarios where an event occurs when an underlying health level
process hits a boundary on one side. Therefore, many of the
examples based on Wiener process reviewed in \citet{lee2006threshold}
are potentially applicable. The Wiener process approach ensures
finiteness of the FHT by a nonzero drift. Our process does so with a
reflecting boundary on the other side for a driftless Wiener
process. When the reflecting boundary is removed, the FHT has a
positive probability of being infinity, which makes it applicable when
a cure rate is needed \citep[Section~5]{lee2006threshold}. Finally,
the DURABLE dataset has additional longitudinally observed blood glucose
levels. To combine these longitudinal observations with the hypoglycemic events
into a joint modeling framework, the transition density of a reflected Brownian
motion would be needed. Incorporating this density into our framework would be
interesting but not trivial. 

\section*{Supplementary Materials}

Supplementary Materials include (A) sensitivity analyses for the application
with different priors and (B) different choices of starting point~$x_0$, 
(C) trace plots for the parameters in independent-frailty model of two
MCMC chains, and (D) code examples.

\appendix
\section{Tail of FHT Density} \label{subsec:asymp}
Here we show that the right tail of the FHT is bounded by an exponential rate.
Note that rates $\lambda_n$  monotonically increase to $\infty$ as $n\to\infty$ 
(so, $\lambda_1$ is the slowest rate),  $c_1>0$, and $|c_n|<1$ for $n\geq 2$. 
It can be shown that density $f(t)$ is asymptotically equivalent to 
$c_1\lambda_1e^{-\lambda_1 t}$ as $t\to\infty$.  

First, let us rewrite the FHT density as follows:
\begin{equation*}
  \begin{split}
    f(t) = \sum_{n=1}^{\infty} c_n \lambda_n e^{-\lambda_n t} 
         = c_1 \lambda_1 e^{-\lambda_1 t} \left( 1 + \frac{1}{c_1\lambda_1}
         \sum_{n=2}^{\infty} c_n\lambda_n e^{-(\lambda_n - \lambda_1)t} \right),
  \end{split}
\end{equation*}
where $\lambda_n = b(2n-1)^2$ and $b = \frac{\sigma^2 \pi^2}{8(\kappa-\nu)^2}$.

Now, consider $h(t) = \sum_{n=2}^{\infty} c_n \lambda_n e^{-(\lambda_n - \lambda_1)t}$ 
and observe that
\begin{equation*}
  \begin{split}
    \lvert h(t)\lvert
    & = \left| \sum_{n=2}^{\infty} c_n \lambda_n
      e^{-(\lambda_n - \lambda_1) t} \right| 
      = e^{-bt} \left| \sum_{n=2}^{\infty} c_n \lambda_n
      e^{-(\lambda_n - \lambda_1- b) t} \right| \\
    & \le e^{-bt} \sum_{n=2}^{\infty} \lvert c_n \lvert
    \lambda_n e^{-(\lambda_n - \lambda_1- b) t} 
      \le e^{-bt} \sum_{n=2}^{\infty}
    \lambda_n e^{-(\lambda_n - \lambda_1- b) t}. 
  \end{split}
\end{equation*}
Therefore, if $\sum_{n=2}^{\infty} \lambda_n e^{-(\lambda_n - \lambda_1- b) t}$
is bounded for all sufficiently large $t$, then $|h(t)|\to 0$ as $t\to\infty$. 
Indeed, since $\lambda_n - \lambda_1 - b = b(2n-1)^2 - 2b > 0$
for $n \ge 2$, for $t > 1$ we have
\begin{equation*}
  \begin{split}
    \sum_{n=2}^{\infty} 
    \lambda_n e^{-(\lambda_n - \lambda_1- b) t}
       \le \sum_{n=2}^{\infty} 
      \lambda_n e^{-(\lambda_n - \lambda_1- b) * 1} 
       =  e^{\lambda_1 + b} \sum_{n=2}^{\infty}
      \lambda_n e^{-\lambda_n} 
       < \infty,
  \end{split}
\end{equation*}
because $\sum_{n=2}^{\infty} \lambda_n e^{-\lambda_n}$ is obviously a convergent series.
Thus, hitting time density $f(t) \sim c_1 \lambda_1
e^{-\lambda_1 t}$ as $t \to \infty$.

This result allows the use of an exponential distribution, with proper scaling,
on the right tail to bound the FHT density as detailed next.

\section{Rejection Sampling Algorithm} \label{subsec:rejSpl}

\begin{figure}[tbp]
\centering
\includegraphics[max width=\textwidth]{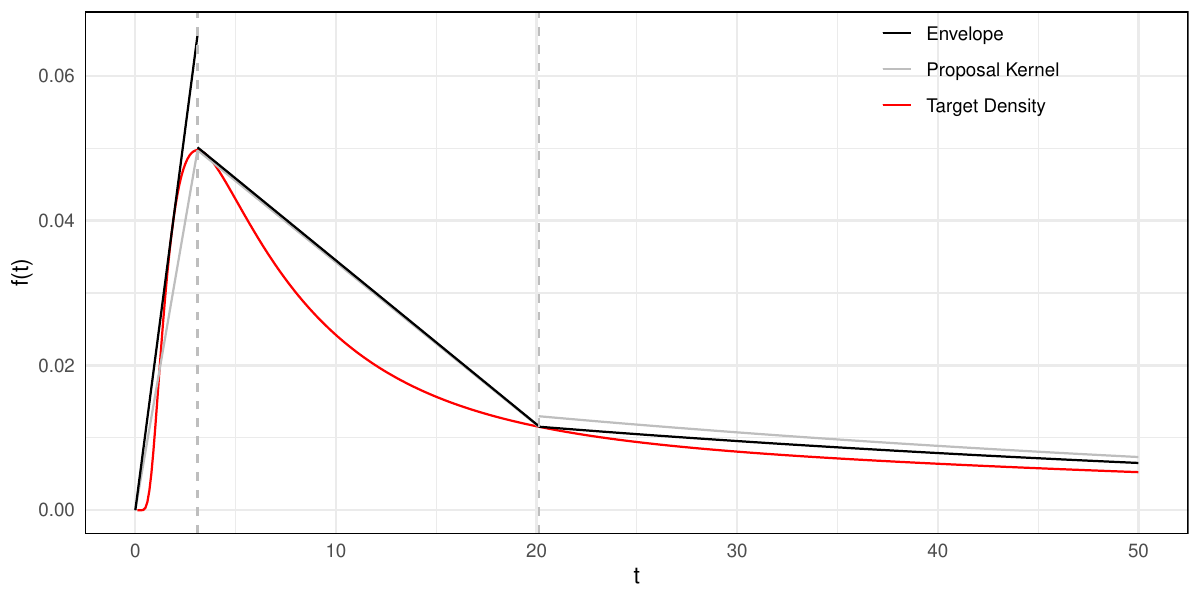}  
\caption{Actual density of the first hitting time distribution of the reflected
Brownian motion with $x_0 = 10$, $\kappa = 20$, $\nu = 3.9$, $\sigma = 2$ is
given as red line. The three-piece kernel and the envelope resulting from the
multiplication of corresponding constants with the kernel are given grey line
and black line respectively. In this plot, $q =0.5$ is considered}
\label{fig:propTarget}
\end{figure}

\begin{algorithm}
  \DontPrintSemicolon
  \SetAlgoLined
  \SetKw{KwAlg}{Algorithm}
  \SetKwProg{KwAlg}{Algorithm}{:}{\KwRet{Y}}
  \KwAlg{}{
    \KwIn{$f$ and $F$, the FHT density and distribution functions; \\
      \phantom{inputs:\,} $q$, a user defined percentile;\\
      \phantom{inputs:\,} $q_{t_m} =  F(t_m)$;\\
      \phantom{inputs:\,} $g_1$, $g_2$, and $g_3$, three component proposals;\\
      \phantom{inputs:\,} $M_1$, $M_2$, and $M_3$, bounding constants for the three components;\\
    }
    \KwOut{$Y$, a draw from target density $f$}

    \BlankLine
    Draw $U \sim \text{Uniform}(0,1)$\;
    \uIf{$U \le q_{t_m}$}{
      \Repeat{$U' \le f(Y)/[M_1 g_1(Y)]$}{
        Draw candidate $Y \sim g_1$\;
        Draw $U' \sim \text{Uniform}(0, 1)$\;
      }
    }
    \uElseIf{$q_{t_m} < U \le q$}{
      \Repeat{$U' \le f(Y)/[M_2 g_2(Y)]$}{
        Draw candidate $Y \sim g_2$\;
        Draw $U' \sim \text{Uniform}(0, 1)$\;
      }
    }
    \uElseIf{$U > q$}{
      \Repeat{$U' \le f(Y)/[M_3 g_3(Y)]$}{
      Draw candidate $Y \sim g_3$\;
      Draw $U' \sim \text{Uniform}(0, 1)$\;
      }
    }

  }
  \caption{Rejection sampling algorithm for drawing one observation from the FHT
    density.} 
  \label{rejSplalgorm}
\end{algorithm}

To sample from the FHT density $f$, we handle the left tail, body, and right
tail separately. Define $g_1(t)$, $g_2(t)$, and $g_3(t)$ as the proposal density
for the left, body, and right components, respectively,
\begin{align*}
  g_1(t) & \propto k_1 t, &t \le t_m, \\
  g_2(t) & \propto k_2 t + f(t_m) - k_2 t_m, & t_m < t \le t_q, \\
  g_3(t) & \propto \exp(-\lambda_1 t), & t > t_q, 
\end{align*}
where $k_1 = f(t_m)/t_m$ and $k_2 = [f(t_m) - f(t_q)]/(t_m - t_q)$.
For illustration, Figure~\ref{fig:propTarget} shows the actual (target) density
of first hitting distribution in red, the kennels of the three proposal
densities in grey, and the envelopes derived by multiplying corresponding
constants to the kernels in black. Given the shape properties of $f$, the $i$th
component of~$f$ can be bounded by $M_i g_i(t)$, where $M_i$ can be identified
by maximizing $f(t) / g_i(t)$ over the domain of $g_i$, $i = 1, 2, 3$.

The rejection sampling algorithm for generating one observation from~$f$ is
given in Algorithm~\ref{rejSplalgorm}.

\bibliographystyle{chicago}
\bibliography{ref}

\begin{thebibliography}{}

\bibitem[\protect\citeauthoryear{Andersen and Gill}{Andersen and
  Gill}{1982}]{andersen1982cox}
Andersen, P.~K. and R.~D. Gill (1982).
\newblock Cox's regression model for counting processes: {A} large sample
  study.
\newblock {\em The Annals of Statistics\/}~{\em 10\/}(4), 1100--1120.

\bibitem[\protect\citeauthoryear{Box-Steffensmeier and
  De~Boef}{Box-Steffensmeier and De~Boef}{2006}]{box2006repeated}
Box-Steffensmeier, J.~M. and S.~De~Boef (2006).
\newblock Repeated events survival models: {T}he conditional frailty model.
\newblock {\em Statistics in Medicine\/}~{\em 25\/}(20), 3518--3533.

\bibitem[\protect\citeauthoryear{Buse, Wolffenbuttel, Herman, Shemonsky, Jiang,
  Fahrbach, Scism-Bacon, and Martin}{Buse et~al.}{2009}]{buse2009durability}
Buse, J.~B., B.~H. Wolffenbuttel, W.~H. Herman, N.~K. Shemonsky, H.~H. Jiang,
  J.~L. Fahrbach, J.~L. Scism-Bacon, and S.~A. Martin (2009).
\newblock {Dura}bility of basal versus lispro mix 75/25 insulin efficacy
  ({Durable}) trial 24-week results: {S}afety and efficacy of insulin lispro
  mix 75/25 versus insulin glargine added to oral antihyperglycemic drugs in
  patients with type 2 diabetes.
\newblock {\em Diabetes Care\/}~{\em 32\/}(6), 1007--1013.

\bibitem[\protect\citeauthoryear{Celeux, Forbes, Robert, and
  Titterington}{Celeux et~al.}{2006}]{celeux2006deviance}
Celeux, G., F.~Forbes, C.~P. Robert, and D.~M. Titterington (2006).
\newblock Deviance information criteria for missing data models.
\newblock {\em Bayesian Analysis\/}~{\em 1\/}(4), 651--673.

\bibitem[\protect\citeauthoryear{{Centers for Disease Control and
  Prevention}}{{Centers for Disease Control and Prevention}}{2022}]{cdc2022}
{Centers for Disease Control and Prevention} (2022).
\newblock National diabetes statistics report website.
\newblock Accessed Oct 28, 2022.

\bibitem[\protect\citeauthoryear{Chang}{Chang}{2004}]{chang2004estimating}
Chang, S.-H. (2004).
\newblock Estimating marginal effects in accelerated failure time models for
  serial sojourn times among repeated events.
\newblock {\em Lifetime Data Analysis\/}~{\em 10\/}(2), 175--190.

\bibitem[\protect\citeauthoryear{Charles-Nelson, Katsahian, and
  Schramm}{Charles-Nelson et~al.}{2019}]{charles2019analyze}
Charles-Nelson, A., S.~Katsahian, and C.~Schramm (2019).
\newblock How to analyze and interpret recurrent events data in the presence of
  a terminal event: {A}n application on readmission after colorectal cancer
  surgery.
\newblock {\em Statistics in Medicine\/}~{\em 38\/}(18), 3476--3502.

\bibitem[\protect\citeauthoryear{Cook and Lawless}{Cook and
  Lawless}{2007}]{cook2007statistical}
Cook, R.~J. and J.~F. Lawless (2007).
\newblock {\em The Statistical Analysis of Recurrent Events}.
\newblock New York: Springer.

\bibitem[\protect\citeauthoryear{Cryer, Axelrod, Grossman, Heller, Montori,
  Seaquist, and Service}{Cryer et~al.}{2009}]{cryer2009evaluation}
Cryer, P.~E., L.~Axelrod, A.~B. Grossman, S.~R. Heller, V.~M. Montori, E.~R.
  Seaquist, and F.~J. Service (2009).
\newblock Evaluation and management of adult hypoglycemic disorders: {A}n
  {E}ndocrine {S}ociety {C}linical {P}ractice {G}uideline.
\newblock {\em The Journal of Clinical Endocrinology and Metabolism\/}~{\em
  94\/}(3), 709--728.

\bibitem[\protect\citeauthoryear{Cryer, Davis, and Shamoon}{Cryer
  et~al.}{2003}]{cryer2003hypoglycemia}
Cryer, P.~E., S.~N. Davis, and H.~Shamoon (2003).
\newblock Hypoglycemia in diabetes.
\newblock {\em Diabetes Care\/}~{\em 26\/}(6), 1902--1912.

\bibitem[\protect\citeauthoryear{{de Valpine}, Turek, Paciorek,
  Anderson-Bergman, {Temple Lang}, and Bodik}{{de Valpine}
  et~al.}{2017}]{devalpine2017programming}
{de Valpine}, P., D.~Turek, C.~Paciorek, C.~Anderson-Bergman, D.~{Temple Lang},
  and R.~Bodik (2017).
\newblock Programming with models: {W}riting statistical algorithms for general
  model structures with {NIMBLE}.
\newblock {\em Journal of Computational and Graphical Statistics\/}~{\em 26},
  403--413.

\bibitem[\protect\citeauthoryear{DeRosa and Cryer}{DeRosa and
  Cryer}{2004}]{derosa2004hypoglycemia}
DeRosa, M.~A. and P.~E. Cryer (2004).
\newblock Hypoglycemia and the sympathoadrenal system: {N}eurogenic symptoms
  are largely the result of sympathetic neural, rather than adrenomedullary,
  activation.
\newblock {\em American Journal of Physiology-Endocrinology and
  Metabolism\/}~{\em 287\/}(1), E32--E41.

\bibitem[\protect\citeauthoryear{Dey, Chen, and Chang}{Dey
  et~al.}{1997}]{dey1997bayesian}
Dey, D.~K., M.-H. Chen, and H.~Chang (1997).
\newblock Bayesian approach for nonlinear random effects models.
\newblock {\em Biometrics\/}~{\em 53\/}(4), 1239--1252.

\bibitem[\protect\citeauthoryear{Doubleday, Zhou, Zhou, and Fu}{Doubleday
  et~al.}{2022}]{doubleday2022risk}
Doubleday, K., J.~Zhou, H.~Zhou, and H.~Fu (2022).
\newblock Risk controlled decision trees and random forests for precision
  medicine.
\newblock {\em Statistics in Medicine\/}~{\em 41\/}(4), 719--735.

\bibitem[\protect\citeauthoryear{Duchateau, Janssen, Kezic, and
  Fortpied}{Duchateau et~al.}{2003}]{duchateau2003evolution}
Duchateau, L., P.~Janssen, I.~Kezic, and C.~Fortpied (2003).
\newblock Evolution of recurrent asthma event rate over time in frailty models.
\newblock {\em Journal of the Royal Statistical Society Series C: Applied
  Statistics\/}~{\em 52\/}(3), 355--363.

\bibitem[\protect\citeauthoryear{Economou, Malefaki, and Caroni}{Economou
  et~al.}{2015}]{economou2015bayesian}
Economou, P., S.~Malefaki, and C.~Caroni (2015).
\newblock Bayesian threshold regression model with random effects for recurrent
  events.
\newblock {\em Methodology and Computing in Applied Probability\/}~{\em
  17\/}(4), 871--898.

\bibitem[\protect\citeauthoryear{Folks and Chhikara}{Folks and
  Chhikara}{1978}]{folks1978inverse}
Folks, J.~L. and R.~S. Chhikara (1978).
\newblock The inverse gaussian distribution and its statistical application ---
  {A} review.
\newblock {\em Journal of the Royal Statistical Society: Series B
  (Methodological)\/}~{\em 40\/}(3), 263--275.

\bibitem[\protect\citeauthoryear{Fu, Luo, and Qu}{Fu
  et~al.}{2016}]{fu2016hypoglycemic}
Fu, H., J.~Luo, and Y.~Qu (2016).
\newblock Hypoglycemic events analysis via recurrent time-to-event ({H}eart)
  models.
\newblock {\em Journal of Biopharmaceutical Statistics\/}~{\em 26\/}(2),
  280--298.

\bibitem[\protect\citeauthoryear{Geisser and Eddy}{Geisser and
  Eddy}{1979}]{geisser1979predictive}
Geisser, S. and W.~F. Eddy (1979).
\newblock A predictive approach to model selection.
\newblock {\em Journal of the American Statistical Association\/}~{\em
  74\/}(365), 153--160.

\bibitem[\protect\citeauthoryear{Gelfand and Dey}{Gelfand and
  Dey}{1994}]{gelfand1994bayesian}
Gelfand, A.~E. and D.~K. Dey (1994).
\newblock {B}ayesian model choice: {A}symptotics and exact calculations.
\newblock {\em Journal of the Royal Statistical Society: Series B
  (Methodological)\/}~{\em 56\/}(3), 501--514.

\bibitem[\protect\citeauthoryear{Gelman}{Gelman}{2006}]{gelman2006prior}
Gelman, A. (2006).
\newblock Prior distributions for variance parameters in hierarchical models
  (comment on article by {B}rowne and {D}raper).
\newblock {\em Bayesian Analysis\/}~{\em 1\/}(3), 515--534.

\bibitem[\protect\citeauthoryear{Heidelberger and Welch}{Heidelberger and
  Welch}{1983}]{heidelberger1983simulation}
Heidelberger, P. and P.~D. Welch (1983).
\newblock Simulation run length control in the presence of an initial
  transient.
\newblock {\em Operations Research\/}~{\em 31\/}(6), 1109--1144.

\bibitem[\protect\citeauthoryear{Hofert, Kojadinovic, M{\"a}chler, and
  Yan}{Hofert et~al.}{2018}]{hofert2018elements}
Hofert, M., I.~Kojadinovic, M.~M{\"a}chler, and J.~Yan (2018).
\newblock {\em Elements of Copula Modeling with R}.
\newblock Springer.

\bibitem[\protect\citeauthoryear{Hu, Wang, and Yang}{Hu
  et~al.}{2012}]{hu2012hitting}
Hu, Q., Y.~Wang, and X.~Yang (2012).
\newblock The hitting time density for a reflected brownian motion.
\newblock {\em Computational Economics\/}~{\em 40\/}(1), 1--18.

\bibitem[\protect\citeauthoryear{Huang and Chen}{Huang and
  Chen}{2003}]{huang2003marginal}
Huang, Y. and Y.~Q. Chen (2003).
\newblock Marginal regression of gaps between recurrent events.
\newblock {\em Lifetime Data Analysis\/}~{\em 9}, 293--303.

\bibitem[\protect\citeauthoryear{Klein}{Klein}{1992}]{klein1992semiparametric}
Klein, J.~P. (1992).
\newblock Semiparametric estimation of random effects using the {C}ox model
  based on the {EM} algorithm.
\newblock {\em Biometrics\/}~{\em 48\/}(3), 795--806.

\bibitem[\protect\citeauthoryear{Lawless and Nadeau}{Lawless and
  Nadeau}{1995}]{lawless1995some}
Lawless, J.~F. and C.~Nadeau (1995).
\newblock Some simple robust methods for the analysis of recurrent events.
\newblock {\em Technometrics\/}~{\em 37\/}(2), 158--168.

\bibitem[\protect\citeauthoryear{Lee, Wei, Amato, and Leurgans}{Lee
  et~al.}{1992}]{lee1992cox}
Lee, E.~W., L.~Wei, D.~A. Amato, and S.~Leurgans (1992).
\newblock {C}ox-type regression analysis for large numbers of small groups of
  correlated failure time observations.
\newblock In J.~P. Klein and P.~K. Goel (Eds.), {\em Survival Analysis: State
  of the Art}, pp.\  237--247. Springer.

\bibitem[\protect\citeauthoryear{Lee}{Lee}{2019}]{lee2019survey}
Lee, M.-L.~T. (2019).
\newblock A survey of threshold regression for time-to-event analysis and
  applications.
\newblock {\em Taiwanese Journal of Mathematics\/}~{\em 23\/}(2), 293--305.

\bibitem[\protect\citeauthoryear{Lee and Whitmore}{Lee and
  Whitmore}{2006}]{lee2006threshold}
Lee, M.-L.~T. and G.~A. Whitmore (2006).
\newblock Threshold regression for survival analysis: {M}odeling event times by
  a stochastic process reaching a boundary.
\newblock {\em Statistical Science\/}~{\em 21\/}(4), 501--513.

\bibitem[\protect\citeauthoryear{Lin, Wei, Yang, and Ying}{Lin
  et~al.}{2000}]{lin2000semiparametric}
Lin, D.~Y., L.-J. Wei, I.~Yang, and Z.~Ying (2000).
\newblock Semiparametric regression for the mean and rate functions of
  recurrent events.
\newblock {\em Journal of the Royal Statistical Society: Series B (Statistical
  Methodology)\/}~{\em 62\/}(4), 711--730.

\bibitem[\protect\citeauthoryear{Luo, Huang, and Wang}{Luo
  et~al.}{2013}]{luo2013quantile}
Luo, X., C.-Y. Huang, and L.~Wang (2013).
\newblock Quantile regression for recurrent gap time data.
\newblock {\em Biometrics\/}~{\em 69\/}(2), 375--385.

\bibitem[\protect\citeauthoryear{Ma, Peng, Huang, and Fu}{Ma
  et~al.}{2021}]{ma2021heterogeneous}
Ma, H., L.~Peng, C.-Y. Huang, and H.~Fu (2021).
\newblock Heterogeneous individual risk modelling of recurrent events.
\newblock {\em Biometrika\/}~{\em 108\/}(1), 183--198.

\bibitem[\protect\citeauthoryear{Maity, Basu, and Ghosh}{Maity
  et~al.}{2021}]{maity2021bayesian}
Maity, A.~K., S.~Basu, and S.~Ghosh (2021).
\newblock {B}ayesian criterion-based variable selection.
\newblock {\em Journal of the Royal Statistical Society Series C: Applied
  Statistics\/}~{\em 70\/}(4), 835--857.

\bibitem[\protect\citeauthoryear{Malefaki, Economou, and Caroni}{Malefaki
  et~al.}{2015}]{malefaki2015modelling}
Malefaki, S., P.~Economou, and C.~Caroni (2015).
\newblock Modelling times between events with a cured fraction using a first
  hitting time regression model with individual random effects.
\newblock In C.~P. Kitsos, T.~A. Oliveira, A.~Rigas, and S.~Gulati (Eds.), {\em
  Theory and Practice of Risk Assessment}, pp.\  45--65. Springer.

\bibitem[\protect\citeauthoryear{Manda and Meyer}{Manda and
  Meyer}{2005}]{manda2005bayesian}
Manda, S.~O. and R.~Meyer (2005).
\newblock {B}ayesian inference for recurrent events data using time-dependent
  frailty.
\newblock {\em Statistics in Medicine\/}~{\em 24\/}(8), 1263--1274.

\bibitem[\protect\citeauthoryear{Pennell, Whitmore, and Ting~Lee}{Pennell
  et~al.}{2010}]{pennell2010bayesian}
Pennell, M.~L., G.~Whitmore, and M.-L. Ting~Lee (2010).
\newblock {B}ayesian random-effects threshold regression with application to
  survival data with nonproportional hazards.
\newblock {\em Biostatistics\/}~{\em 11\/}(1), 111--126.

\bibitem[\protect\citeauthoryear{Plummer, Best, Cowles, and Vines}{Plummer
  et~al.}{2006}]{plummer2006coda}
Plummer, M., N.~Best, K.~Cowles, and K.~Vines (2006).
\newblock {CODA}: {C}onvergence diagnosis and output analysis for {MCMC}.
\newblock {\em R News\/}~{\em 6\/}(1), 7--11.

\bibitem[\protect\citeauthoryear{Prentice, Williams, and Peterson}{Prentice
  et~al.}{1981}]{prentice1981regression}
Prentice, R.~L., B.~J. Williams, and A.~V. Peterson (1981).
\newblock On the regression analysis of multivariate failure time data.
\newblock {\em Biometrika\/}~{\em 68\/}(2), 373--379.

\bibitem[\protect\citeauthoryear{Schaubel and Cai}{Schaubel and
  Cai}{2004}]{schaubel2004regression}
Schaubel, D.~E. and J.~Cai (2004).
\newblock Regression methods for gap time hazard functions of sequentially
  ordered multivariate failure time data.
\newblock {\em Biometrika\/}~{\em 91\/}(2), 291--303.

\bibitem[\protect\citeauthoryear{Schr{\"o}dinger}{Schr{\"o}dinger}{1915}]{schrodinger1915theorie}
Schr{\"o}dinger, E. (1915).
\newblock Zur theorie der fall-und steigversuche an teilchen mit brownscher
  bewegung.
\newblock {\em Physikalische Zeitschrift\/}~{\em 16}, 289--295.

\bibitem[\protect\citeauthoryear{Seaquist, Anderson, Childs, Cryer,
  Dagogo-Jack, Fish, Heller, Rodriguez, Rosenzweig, and Vigersky}{Seaquist
  et~al.}{2013}]{seaquist2013hypoglycemia}
Seaquist, E.~R., J.~Anderson, B.~Childs, P.~Cryer, S.~Dagogo-Jack, L.~Fish,
  S.~R. Heller, H.~Rodriguez, J.~Rosenzweig, and R.~Vigersky (2013).
\newblock Hypoglycemia and diabetes: {A} report of a workgroup of the
  {A}merican {D}iabetes {A}ssociation and the {E}ndocrine {S}ociety.
\newblock {\em Diabetes Care\/}~{\em 36\/}(5), 1384--1395.

\bibitem[\protect\citeauthoryear{Spiegelhalter, Best, Carlin, and Van
  Der~Linde}{Spiegelhalter et~al.}{2002}]{spiegelhalter2002bayesian}
Spiegelhalter, D.~J., N.~G. Best, B.~P. Carlin, and A.~Van Der~Linde (2002).
\newblock {B}ayesian measures of model complexity and fit.
\newblock {\em Journal of the Royal Statistical Society: Series B (Statistical
  Methodology)\/}~{\em 64\/}(4), 583--639.

\bibitem[\protect\citeauthoryear{Sun, Park, and Sun}{Sun
  et~al.}{2006}]{sun2006additive}
Sun, L., D.-H. Park, and J.~Sun (2006).
\newblock The additive hazards model for recurrent gap times.
\newblock {\em Statistica Sinica\/}~{\em 16\/}(3), 919--932.

\bibitem[\protect\citeauthoryear{Towler, Havlin, Craft, and Cryer}{Towler
  et~al.}{1993}]{towler1993mechanism}
Towler, D.~A., C.~E. Havlin, S.~Craft, and P.~Cryer (1993).
\newblock Mechanism of awareness of hypoglycemia: {P}erception of neurogenic
  (predominantly cholinergic) rather than neuroglycopenic symptoms.
\newblock {\em Diabetes\/}~{\em 42\/}(12), 1791--1798.

\bibitem[\protect\citeauthoryear{Wei, Lin, and Weissfeld}{Wei
  et~al.}{1989}]{wei1989regression}
Wei, L.-J., D.~Y. Lin, and L.~Weissfeld (1989).
\newblock Regression analysis of multivariate incomplete failure time data by
  modeling marginal distributions.
\newblock {\em Journal of the American Statistical Association\/}~{\em
  84\/}(408), 1065--1073.

\bibitem[\protect\citeauthoryear{Whitmore, Ramsay, and Aaron}{Whitmore
  et~al.}{2012}]{whitmore2012recurrent}
Whitmore, G., T.~Ramsay, and S.~Aaron (2012).
\newblock Recurrent first hitting times in {W}iener diffusion under several
  observation schemes.
\newblock {\em Lifetime Data Analysis\/}~{\em 18\/}(2), 157--176.

\bibitem[\protect\citeauthoryear{Wild, von Maltzahn, Brohan, Christensen,
  Clauson, and Gonder-Frederick}{Wild et~al.}{2007}]{wild2007critical}
Wild, D., R.~von Maltzahn, E.~Brohan, T.~Christensen, P.~Clauson, and
  L.~Gonder-Frederick (2007).
\newblock A critical review of the literature on fear of hypoglycemia in
  diabetes: {I}mplications for diabetes management and patient education.
\newblock {\em Patient Education and Counseling\/}~{\em 68\/}(1), 10--15.

\bibitem[\protect\citeauthoryear{Xu, Chiou, Yan, Marr, and Huang}{Xu
  et~al.}{2020}]{xu2020generalized}
Xu, G., S.~H. Chiou, J.~Yan, K.~Marr, and C.-Y. Huang (2020).
\newblock Generalized scale-change models for recurrent event processes under
  informative censoring.
\newblock {\em Statistica Sinica\/}~{\em 30}, 1773.

\end{thebibliography}

\end{document}